\documentclass[preprint]{elsarticle}

\usepackage{amssymb}
\usepackage{amsmath}
\usepackage{amsthm}
\usepackage{mathtools}
\usepackage{natbib}

\usepackage{float}
\usepackage{caption} 
\usepackage{array}
\usepackage{framed}
\usepackage{dsfont}
\usepackage{xcolor}
\usepackage{subcaption}
\usepackage{hyperref}
\usepackage{booktabs}
\usepackage{multirow}

\journal{Journal of the Korean Statistical Society}

\begin{document}

\begin{frontmatter}

\title{Short Communication: Detecting Possibly Frequent Change-points: Wild Binary Segmentation 2 and Steepest-drop Model Selection}

\author{
{\sc Robert Lund} \\
Department of Statistics \\
The University of California - Santa Cruz \\
Santa Cruz, CA 95064 \\
and \\
{\sc Xueheng Shi} \\
School of Mathematics and Statistics \\
Clemson University \\
Clemson, SC 29634-0975}

\begin{abstract}
This article comments on the new version of wild binary segmentation in \cite{fryzlewicz2014wbs}.
\end{abstract}


\end{frontmatter}


\section{Overview}

We thank the author and journal for the opportunity to comment on the newest version of wild binary segmentation (WBS).  The paper was well written, with the major algorithmic ideas expertly articulated --- it was an interesting read.

Multiple changepoint problems have received considerable attention over the last two decades.  While the community has seemingly converged on single changepoint methods, concensus on how to tackle the multiple changepoint case is still lacking. Two major camps have seemingly evolved.  

First, the penalized likelihood camp minimizes a penalized likelihood objective function that selects the number of changepoints and other model parameters.  These methods are computationally intense as there are $2^{T-1}$ different admissible changepoint configurations to be explored in a sequence of length $T$ (time one cannot be a changepoint).  An exhaustive model search evaluating all changepoint configurations is not possible with $T=100$ for some simple models. Some authors \cite{Davis_etal_2006, Li_Lund_2012} use a genetic algorithm, which is an intelligent random walk search, to optimize the penalized likelihood.  Unfortunately, such a randomized search may fail to identify a global minimum; moreover, one may need to wait several days for the genetic algorithm to run when $T=1000$.  While penalized likelihoods work in a variety of scenarios, including time series errors, the camp argues about what penalty is optimal.

The second major camp takes a more algorithmic approach, attempting to devise fast estimation routines that perform reasonably.  The algorithmic approach adopted in this paper is an improved version of WBS, with a new technique used to select the number of changepoints that is dubbed a steepest-drop to low levels (SDLL) criteria.  Binary segmentation, the earliest multiple changepoint algorithm, applies single changepoint techniques to identify the most prominent changepoint, and then splits the series into two subsegments about the flagged changepoint time (should it exist). The process is repeated iteratively to all subsegments until they test as changepoint free.  Binary segmentation is computationally fast and conceptually simple, but its nature as a greedy algorithm can leave it fooled.  WBS \cite{fryzlewicz2014wbs} and narrowest-over-threshold detection \cite{baranowski2019} methods overcome binary segmentation weaknesses by drawing many random subintervals (subsegments), hoping that a few subsegments will contain one and only one changepoint (this improves estimation).  The injected randomness enables binary segmentation to escape a local optimum.

The improvements to the original WBS algorithm made here are two-fold.   The first simply samples the random subsegments in a data-driven way at stages in lieu of generating all subsegments at the algorithm's onset.  When subsegments become small enough, all further subsegments are exhaustively considered. This procedure yields candidate estimated changepoint configurations with $0, 1, \ldots , T-1$ changepoints and is reasonably fast.  The second improvement replaces WBS thresholding methods to select the number of changepoints to a SDLL criteria that is shown to have a theoretical basis.  The algorithm is dubbed WBS2-SDLL.  Many of our comments below apply both to WBS and WBS2-SDLL.

\section{Distances between changepoint configurations}

To compare different methods in both the frequent and infrequent changepoint cases, a distance between two changepoint configurations, denoted by $\mathcal{C}_1 = (m; \tau_1, \ldots, \tau_m)$ (with $m$ changepoints) and $\mathcal{C}_2 = (k; \eta_1, \ldots, \eta_k)$ (with $k$ changepoints), is needed.   We will use the distance
\begin{equation*}      
    d(\mathcal{C}_1, \mathcal{C}_2)=|m-k| +
    \min \mathcal{A}(\mathcal{C}_1, \mathcal{C}_2).
\end{equation*}
discussed in \cite{Shi_etal_2020}.  The term $|m-k|$ accounts for the discrepancy between the changepoint counts in the two configurations and the "minimum component" $\min \mathcal{A}(\mathcal{C}_1, \mathcal{C}_2)$ measures how well the two changepoint time sets align with one and other.  This term is computed via the linear assignment structure
\begin{equation*}
\mathcal{A}(\mathcal{C}_1, \mathcal{C}_2)
=
\sum_{i=1}^k\sum_{j=1}^m c_{ij}x_{ij}
\end{equation*}
subject to the constraints $\sum_{i=1}^kx_{ij}=1$ for $j=1, \cdots,m$, $\sum_{j=1}^mx_{ij} \leq 1$ for $i=1, \cdots, k$, with $x_{ij} \in \{0,1\}$.   Think of $c_{ij}$ as the cost of assigning $\tau_i$ to $\eta_j$ in the matching; the particular cost used here is $c_{ij} = |\tau_i-\eta_j|/N$.   Finally, $x_{ij}$ is a decision variable satisfying
\begin{equation*}
x_{ij} = 
    \begin{cases}
    \; 1 \qquad &\text{if $\tau_i$ is assigned to $\eta_j$}\\
    \; 0 \qquad &\text{otherwise}
    \end{cases}.
\end{equation*}
Efficient algorithms are available to rapidly compute this quantity \cite{burkard2012assignment}. The minimum term can be shown to be bounded by unity. In the case of frequent changepoints, the $|m-k|$ term dominates. This distance will be used below to compare true and estimated changepoint configurations. 

\section{Frequent Changepoint Settings}. Beginning our individual comments, the pursuit of the high-frequency changepoint case seems somewhat over-emphasized. Indeed, with regard to the extreme teeth signal in Figure 1, a time series analyst would be remissed if their exploratory model/analysis neglected a seasonal (periodic) component.  A power spectrum should readily identify the period of the data and standard time series regression techniques would estimate the periodic mean and noise structure.  One would obtain a more parsimonious model.

Of course, the extreme teeth signal could be replaced by one where the teeth have varying widths and heights to negate the above complaint, but in this case, other statistical techniques are available.  Specifically, if one were given a time series with such structure, a non-parametric regression analysis for the mean would be our first urge --- especially in the absence of physical justification for the mean shifts.  Phrased another way, an example where frequent mean shifts are physically plausible would be appreciated.   The fitted changepoint configuration in the London house price series in Figure 7 did not excite us: some of these changepoints seem more attributable to the positive correlations found in economic series than to mean shifts.

For full disclosure, our interests in the multiple changepoint problem lie with climate time series, where weather stations have their gauges changed or are physically moved an average of six times per century in the United States \cite{Mitchell1953}. This scenario dictates the need for many abrupt changes in the mean. And because weather is correlated, one must allow for correlation in the model errors, something absent here (more on this below).

The above said, high dimensional economic series may indeed have frequent changes in some of its component series due to corporate mergers, political instability, etc. (imagine daily tracking of the 500 individual stocks in the S\&P 500 stock index).  As such, our concern is primarily confined to the univariate setting.

\section{WBS Performance in Infrequent Changepoint Settings}

Some of the claims that WBS methods work well in infrequent changepoint settings did not jibe with our investigations of its performance.  Elaborating, our classical task typically involves the homogenization of a century of annually average temperatures at a station ($T=100$).  With this $T$, one typically has about three true shifts for United States series (that is, roughly half of the gauge change and station move times induce a true mean shift), but sometimes there are none.

As such, a simulation was done on a time series with no changepoints:  simple IID $N(0,1)$ noise with lengths $T=100$ and $T=500$. Here, WBS and WBS2-SDLL were compared to two often-used penalized likelihood methods:  BIC and mBIC \cite{zhang2007mbic}. The WBS and WBS2-SDLL methods were executed via the author's packages, which use a CUSUM statistic to assess individual subsegments.  A genetic algorithm from R was used to optimize the penalized likelihoods with an unknown variance parameter $\sigma^2$  \cite{scrucca2013ga}. The results are summarized in the table below of average distances and empirical probabilities of getting one or more changepoints. Each table entry aggregates 1000 independent simulations.

\begin{center}
\begin{table}[ht]
 \caption{Average False Positive Rates and Distances}
  \begin{tabular}{lcccc}
    \toprule
    \multirow{2}{*}{Methods} &
      \multicolumn{2}{c}{T=100} &
      \multicolumn{2}{c}{T=500} \\
      & {False Positive} & {Distance} & {False Positive} & {Distance}  \\
      \midrule
BIC	&0.003	&0.003	&0.000	&0.000\\
mBIC	&0.034	&0.040	&0.006	&0.006\\
WBS	&0.204	&0.390	&0.079	&0.103\\
WBS-SDLL	&0.164	&1.183	&0.133	&0.352\\
\bottomrule
\end{tabular}
\end{table}
\end{center}

For a null hypothesis of no changepoints, a false positive rate of around 20 percent is not particularly good.  While WBS2-SDLL works better than WBS, performance issues remain.   We refer to \cite{Shi_etal_2020} for additional comparisons of multiple changepoint techniques.  Our overarching point is that a more extensive comparison of these techniques is needed in low- and mid-frequency changepoint settings.  We would not care about performance in high-frequency settings should performance in low- and mid-frequency settings need to be sacrificed.

\section{Threshold Selection, Asymptotics, and Consistency}

The poor performance of WBS and WBS2-SDLL above is likely still traced to threshold selection issues (or their equivalents).  Towards this, additional rigorous probabilistic justification seems needed.  Even in the original WBS setting, this does not appear easy.  Elaborating, a CUSUM changepoint statistic for Gaussian data computed over all admissible changepoint times in the subsegment $[a,b]$ is distributed as the maximum absolute value sampled from a Gaussian process with a particular covariance structure, often asymptotically expressed via supremums of Brownian bridges --- see \cite{Robbins_etal_JASA} for a recent treatment.  The foundational idea of WBS is to randomly draw many subsegments $[a, b]$, uniformly distributed over the observation times $\{ 1, 2, \ldots , T \}$.  When one takes a maximum over the maximums from the many different subsegments, extreme value distributions would, in principle, arise.   But it is not clear how to proceed as correlation in the individual subsegment maximums would exist should the corresponding intervals overlap; moreover, the subsegment maximums would depend on $a$ and $b$, and there would be many short intervals to deal with --- the Brownian bridge approximation might be off. 

As such, it is not clear how to proceed with WBS threshold selection on technical grounds.  Likewise, the SDLL methods here sound interesting, but we had issues following the theoretical proofs.   The claim that \cite{wang2020univariate} has fixed all theoretical aspects of WBS was not appreciated here; indeed, we have ground zero confusion with the asymptotic setup presented both here and in \cite{wang2020univariate}.   Elaborating, it would seem here that as $T \rightarrow \infty$, the individual changepoint times $\eta_i$ need to depend on $T$ for this scenario to make sense.  But this is not stated; moreover, in such a scenario, we would need $\eta_i$ to converge in some sense as $T \rightarrow \infty$.  Consistently estimating all changepoint times in the traditional statistical sense --- their mean squared errors go to zero --- seems impossible unless an infinity of observations is taken between all changepoint times. Related here:  why do the mean shift sizes in Assumption 3.1 (d) depend on $T$?  This seems unnecessary.  Ditto Assumption 3.1 (c) with a finite number of changepoints.  Clarification:  we believe that the number of changepoints $N$ is not changing in $T$, but this is not clear either.  Infill asymptotics for multiple changepoint problems are considered in \cite{Davis_etal_2006, li2019multiple}: the proofs are long and hard.  In particular, we would appreciate discourse that illuminated all assumptions and proof steps, showing where normality is invoked, what maximal inequalities are used, etc.

\section{Correlation}

An assumption made here is that the data are IID and normally distributed.  Independence is often questionable for time series data; moreover, neglecting correlation can drastically influence estimated changepoint configurations.  As remarked above, the changes in Figure 7 seem just as attributable to correlation than to true mean shifts.

In this paper, if the variance parameter $\sigma^2$ were known, what would stop us from using PELT, FPOP, or another related dynamic programming technique with a penalized likelihood for rapid computation?  A crucial component here seemingly lies with estimating $\sigma^2$ accurately at the onset in the possible presence of many mean shifts.  The author mentions MOSUM methods and median absolute deviations of differences.  For time series data, accurately estimating an underlying autocovariance function (say applying to all series segments) seems of paramount importance.  Towards this, the methods here would break down if the series' autocovariance function also changes at every changepoint time --- as in the AR(1) segmentation literature \cite{chakar2017robust}.

\section{Conclusions}

As said in the introduction, the multiple changepoint literature is still in its infancy, and methods that are computationally rapid and produce great segmentations across a robust variety of data assumptions with appropriate technical justification have yet to be developed.  The next decade will no doubt see much more work in this vein; this said, we believe that estimating the correct number of changepoints will always be a tough task (as with most statistical smoothing problems).

The paper here may influence our work.  WBS methods are known to be aggressive in that they tend to overestimate the number of changepoints.  One could take the estimated changepoint configuration here and tune it with some penalized likelihood method.  Elaborating, if a series of length 10,000 could be reduced to say 100 good candidate changepoint times to further scrutinize, a genetic algorithm would make quick work of the task.  Such an initial configuration could also be used to put a prior on the changepoint times in the configuration; see \cite{li2019multiple} for such methods.

\bibliographystyle{plain}
\bibliography{mybibfile}

\begin{thebibliography}{10}

\bibitem{baranowski2019}
R.~Baranowski, Y.~Chen, and P.~Fryzlewicz.
\newblock Narrowest-over-threshold detection of multiple change points and
  change-point-like features.
\newblock {\em Journal of the Royal Statistical Society: Series B},
  81(3):649--672, 2019.

\bibitem{burkard2012assignment}
R.~Burkard, M.~Dell'Amico, and S.~Martello.
\newblock {\em Assignment Problems}, volume 106.
\newblock Society of Industrial and Applied Mathematics, revised edition, 2012.

\bibitem{chakar2017robust}
S.~Chakar, E.~Lebarbier, C.~L{\'e}vy-Leduc, and S.~Robin.
\newblock A robust approach for estimating change-points in the mean of an
  $\text{AR}(1)$ process.
\newblock {\em Bernoulli}, 23(2):1408--1447, 2017.

\bibitem{Davis_etal_2006}
R.~A. Davis, T.~C.~M. Lee, and G.~A. Rodriguez-Yam.
\newblock Structural break estimation for nonstationary time series models.
\newblock {\em Journal of the American Statistical Association},
  101(473):223--239, 2006.

\bibitem{fryzlewicz2014wbs}
P.~Fryzlewicz.
\newblock Wild binary segmentation for multiple change-point detection.
\newblock {\em The Annals of Statistics}, 42(6):2243--2281, 2014.

\bibitem{Li_Lund_2012}
S.~Li and R.~B. Lund.
\newblock Multiple changepoint detection via genetic algorithms.
\newblock {\em Journal of Climate}, 25(2):674--686, 2012.

\bibitem{li2019multiple}
Y.~Li, R.~B. Lund, and A.~Hewaarachchi.
\newblock Multiple changepoint detection with partial information on
  changepoint times.
\newblock {\em Electronic Journal of Statistics}, 13(2):2462--2520, 2019.

\bibitem{Mitchell1953}
J.~M. Mitchell~Jr.
\newblock On the causes of instrumentally observed secular temperature trends.
\newblock {\em Journal of Meteorology}, 10(4):244--261, 1953.

\bibitem{Robbins_etal_JASA}
M.~W. Robbins, C.~M. Gallagher, and R.~B. Lund.
\newblock A general regression changepoint test for time series data.
\newblock {\em Journal of the American Statistical Association},
  111(514):670--683, 2016.

\bibitem{scrucca2013ga}
L.~Scrucca.
\newblock {GA}: a package for genetic algorithms in {R}.
\newblock {\em Journal of Statistical Software}, 53(4):1--37, 2013.

\bibitem{Shi_etal_2020}
X.~Shi, C.~M. Gallagher, R.~B. Lund, and R.~Killick.
\newblock A statistical comparison of single and multiple changepoint
  techniques for time series data.
\newblock {\em In Preparation}, 2020.

\bibitem{wang2020univariate}
D.~Wang, Y.~Yu, and A.~Rinaldo.
\newblock Univariate mean change point detection: Penalization, {CUSUM} and
  optimality.
\newblock {\em Electronic Journal of Statistics}, 14(1):1917--1961, 2020.

\bibitem{zhang2007mbic}
N.~R. Zhang and D.~O. Siegmund.
\newblock A modified {B}ayes information criterion with applications to the
  analysis of comparative genomic hybridization data.
\newblock {\em Biometrics}, 63(1):22--32, 2007.

\end{thebibliography}

\end{document}